\documentclass[twocolumn,prl,aps]{revtex4}
\usepackage{amsmath}
\usepackage{amssymb}
\usepackage{amsfonts}
\usepackage{graphicx}

\begin{document}
\title
{Fine Structure of Cyclotron Resonance in a Two-Dimensional Electron System}

\author{V.~M.~Muravev, I.~V.~Andreev, S.~I.~Gubarev, V.~N.~Belyanin, I.~V.~Kukushkin}
\affiliation{Institute of Solid State Physics, RAS, Chernogolovka 142432, Russia}
\date{\today}
\date{\today}

\begin{abstract}
It is established that cyclotron resonance (CR) in a high-quality GaAs/AlGaAs two-dimensional electron system (2DES) originates as a \textit{pure} resonance, that does not hybridize with dimensional magnetoplasma excitations. The magnetoplasma resonances form a fine structure of the CR. The observed fine structure of the CR results from the interplay between coherent radiative and incoherent collisional mechanisms of 2D plasma relaxation. We show that the range of 2DES filling factors from which the phenomenon arises is intimately connected to the fundamental fine-structure constant. 




\end{abstract}

\maketitle

Cyclotron resonance (CR) spectroscopy is the most direct and convenient method of characterizing the Fermi surface and determining the effective mass of semiconductors~\cite{CR1, CR2, CR3}. In a two-dimensional electron system (2DES), a perpendicular magnetic field quenches the in-plane motion of electrons into cyclotron orbits. In turn, if the phase and polarization of the incident electromagnetic radiation are synchronized with the electron orbital motion, CR is triggered. The first observation of CR in 2DESs was reported for charged carriers in an inversion layer on Si~\cite{Abstreiter:74, Allen:74}. Subsequently, CR spectroscopy has been successfully applied to research on many two-dimensional systems, for example, isotropic 2D carriers in GaAs heterostructures~\cite{Tsui:83, Platzman:84, Zudov:01}, composite fermions~\cite{Kukushkin:02}, heavy fermions in MgZnO/ZnO heterojunctions~\cite{ZnO, Vankov:15}, and anisotropic heavy fermions in AlAs quantum wells~\cite{Shayegan:93, Muravev:05}.   

It is widely believed that the cyclotron resonance originates from the dimensional magnetoplasma resonance~\cite{Allen:83, Heitmann:88, Shikin:92}. The frequency of the hybrid cyclotron magnetoplasma mode is described by the equation
\begin{equation}
\omega^2=\omega_p^2 + \omega_c^2,
\label{f1}
\end{equation} 
where $\omega_c=eB/m^{\ast} c$ is the cyclotron frequency, and $\omega_p$ is the dimensional plasmon frequency~\cite{Stern} (Gaussian units are used in this Letter unless
otherwise stated):
\begin{equation}
\omega_p^2=\frac{2 \pi n_s e^2}{m^{\ast} \varepsilon} q.
\label{f2}
\end{equation}
Here, $n_s$ and $m^{\ast}$ are the density and the effective mass of the 2D electrons, and $\varepsilon$ is the effective permittivity of the surrounding medium. For a narrow 2DES stripe of width $W$, the plasmon wavevector can be approximately described by $q=\pi N/W$ ($N=1, 2, \ldots$ is the number of the plasmon harmonic). 

Contrary to general believe, we discovered that the CR originates as a \textit{pure} resonance, that does not hybridize with dimensional magnetoplasma excitations. The CR rises as a single peak with superimposed contribution from different dimensional magnetoplasma modes. Therefore, initially the CR line shape exhibits a multipeak structure. We show that the CR fine structure is resolved when the coherent radiative 2D plasma relaxation dominates the incoherent collisional damping. In large samples, this regime appears when the 2D conductivity $2 \pi \sigma_{xx}$
is much larger than $c$. Moreover, the range of 2DES filling factors $\nu$ in which the CR fine structure necessarily arises is dictated by the relation $2 \pi \sigma_{xy} > c$. The latter can be rewritten very elegantly as $\nu > 1/\alpha$, where $\alpha = e^2/\hbar c \approx 1/137$ is the fundamental fine-structure constant.

Experiments were performed on a set of high-quality GaAs/AlGaAs single quantum wells with a well width of $25$~nm located approximately $465$~nm below the sample surface. The electron density $n_s$ in different samples ranged from $0{.}5$ to $6{.}6 \times 10^{11}/{\rm cm}^2$, with a typical transport mobility of $5 \times 10^6~{\rm cm}^{2}/{\rm V} \cdot {\rm s}$ ($T = 1.5$~K). Disk-shaped mesas with diameters of $d=1$ and $2.5$~mm were fabricated from these heterostructures. No extra metallic parts were present on top of the mesas or near them. Microwave radiation with frequencies of $10$ to $250$~GHz was supplied to the sample via a rectangular waveguide. The frequency ranges of $10 - 40$~GHz and $40 - 250$~GHz were covered by an Agilent generator and a set of backward wave oscillators, respectively. To detect the plasma resonances, a differential optical technique was used~\cite{Ashkinadze, Gubarev} (see Supplemental Material I and II). We measured the differential luminescence spectrum in the presence and absence of microwave radiation. The integral of the absolute value of the differential spectrum served as a measure of the microwave absorption intensity of the sample. Luminescence spectra were recorded using a charge-coupled device (CCD) camera and a double-grating spectrometer with a spectral resolution of $0.03$~meV. Then the magnitude of the differential signal was studied as a function of the microwave frequency and magnetic field. A stabilized semiconductor laser operating at a wavelength of $780$~nm and a power level of approximately $0{.}1$~mW served as the photoexcitation source. Light from the laser was guided to the sample through an optical quartz fiber inserted inside the waveguide section. The sample was immersed in a liquid helium cryostat with a superconducting magnet. Experiments were performed at a sample temperature of $T = 1.5$~K.

\begin{figure}[!t]
\includegraphics[width=0.47 \textwidth]{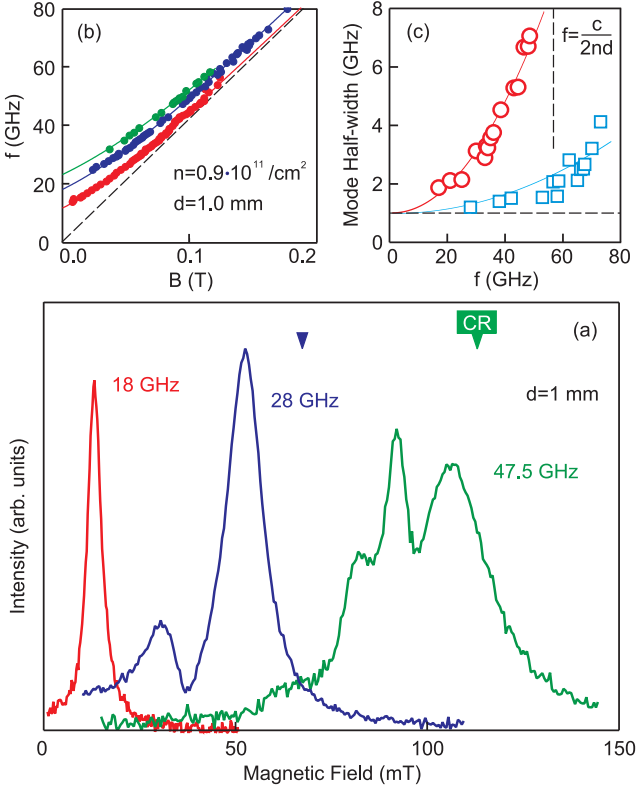}
\caption{(a) Microwave absorption intensity in $d=1$~mm size disk as a function of magnetic field for microwave frequencies $f$ of $18, 28$~and~$47.5$~GHz. Electron density in the sample is $n_s=0{.}9 \times 10^{11}/{\rm cm}^2$. (b) Magnetodispersion of the first three magnetoplasma modes in the sample. Dashed line corresponds to cyclotron frequency in GaAs. (c) Mode linewidth versus resonance frequency for the first $N=1$ (circles) and the second $N=2$ (squares) magnetoplasma modes.}
\label{1}
\end{figure}

Figure~1(a) shows typical resonant microwave absorption curves as a function of the magnetic field measured for several radiation frequencies. The horizontal axis is located at the signal level at which no microwave radiation is supplied to the sample. The data were obtained at an electron density of $0{.}9 \times 10^{11}/{\rm cm}^2$ for a single disk having a diameter of $d=1$~mm. The absorption has the shape of a resonant peak that shifts to higher magnetic field values as the microwave frequency $f$ increases. Second and third peaks emerge at frequencies above $20$~GHz and $30$~GHz, respectively. The physical origin of the observed resonances is best identified by plotting their magnetic field positions versus the incident microwave frequency, as shown in Fig.~1(b). The experimental points follow three lines, which in the strong magnetic field limit tend asymptotically to the cyclotron frequency $\omega_{c}=eB/m^{\ast} c$ ($m^{\ast}=0{.}067 m_0$). Such behavior is characteristic of the dimensional 2D magnetoplasma modes, the magnetodispersions of which are described by equation~(2S) (see Supplemental Material III) with frequencies at zero magnetic field of $f_p=12$, $17$, and $23$~GHz (solid lines in Fig.~1(b))~\cite{Allen:83, Vasiliadou:93}. One interesting point with respect to Fig.~1(a) is that the fundamental ($N=1$) magnetoplasma mode undergoes strong broadening as the radiation frequency increases. At frequencies of $f>40$~GHz, the magnetic field position of the fundamental mode is almost equal to the CR value, $B_c=2 \pi m^{\ast}c f/e$ (arrows in Fig.~1(a)). At the same time, the second and third magnetoplasma harmonics are superimposed on the broad contour of the fundamental mode, which is already indistinguishable from the cyclotron resonance. The observed behavior of the dimensional magnetoplasma modes is a precursor of the appearance of \textit{pure} CR. This effect is thoroughly studied in the remainder of this manuscript.   


Broadening and interplay of different magnetoplasma resonances are at the heart of CR observation. Next, we will study in more detail the broadening of the magnetoplasma excitations. Figure~1(c) shows the dependence of the magnetoplasma resonance half-width on the resonance frequency for the first two modes. Circles and squares correspond to the fundamental magnetoplasma mode ($N=1$) and its second harmonic ($N=2$), respectively. The half-width of the resonance, $\Delta f$, was calculated by multiplying the half-width $\Delta B$ on the magnetic field by the magnetodispersion slope at the field at which the resonance occurs: $\Delta f = (\partial f/\partial B) \Delta B$. After $40$~GHz the second and third magnetoplasma harmonics start to run into the fundamental mode. In this limit, we determine the fundamental mode half-width from the B-field difference between the plasmon resonance maximum and half maximum on the resonance high magnetic field side. To determine the second harmonic width we subtract resonant fundamental mode background from the absorption curve. The magnetoplasma mode width saturates at a constant level of $\Delta f = 1$~GHz in the low-frequency limit. We attribute the relaxation of the plasma mode in this limit to incoherent scattering of charged carriers in the 2DES, $\Delta\omega=\gamma=1/\tau$~\cite{Andreev:14}, where $\gamma$ is the incoherent collisional scattering rate in the sample. As the magnetoplasmon frequency is increased to $c /n f \sim d$, the mode width rises rapidly. One may attribute such broadening of the plasma wave to the superradiant decay $\Gamma$ from the coherent dipole reradiation by oscillating 2D electrons~\cite{Quinn:76, Mikhailov:96, Matov:96, Mikhailov:04, Zudov:14}: $\Delta \omega = \gamma + \Gamma$
\begin{equation}
\Gamma  = \frac{2 \pi \sigma_{xx}}{n c} \times \gamma = \frac{4 \pi n_s e^2}{m^{\ast} (1+ n_{\rm GaAs}) c},
\label{gamma}
\end{equation}
where $n = (1+ n_{\rm GaAs})/2$ is the effective refractive index of the surrounding medium. However, the half-width $\Delta \omega = \gamma + \Gamma$ ($\Delta \omega/2 \pi \approx 10$~GHz for the structure under study) in Eq.~(\ref{gamma}) is not frequency-dependent and greatly exceeds experimentally observed linewidth (Fig.~1(c)). Therefore, the prior view of 2D plasma relaxation requires thorough revision. Indeed, the observed rapid linewidth growth could be easily explained 
in light of the fact that formula~(\ref{gamma}) applies only to a 2DES of infinite size. The coherent superradiant decay of a 2D plasma in a sample of finite size ($d<\lambda$) is given by the corrected equation~\cite{Leavitt:86}     
\begin{equation}
\Gamma_{\rm disk} = \frac{n^3 d^3 \omega_p^4}{9 \pi c^3} \sim \Gamma \times \frac{d^2}{\lambda^2}.
\label{gamma2}
\end{equation}
Indeed, the experimental data in Fig.~1(c) obey the quadratic prediction on the frequency in Eq.~\ref{gamma2} (solid lines). From our experimental data, it follows that the key parameter that separates the regimes of incoherent and coherent 2D plasma relaxation is the commensurability between the radiation wavelength and 2DES disk diameter (Fig.~1(c), vertical dashed line).

\begin{figure}[!t]
\includegraphics[width=0.47 \textwidth]{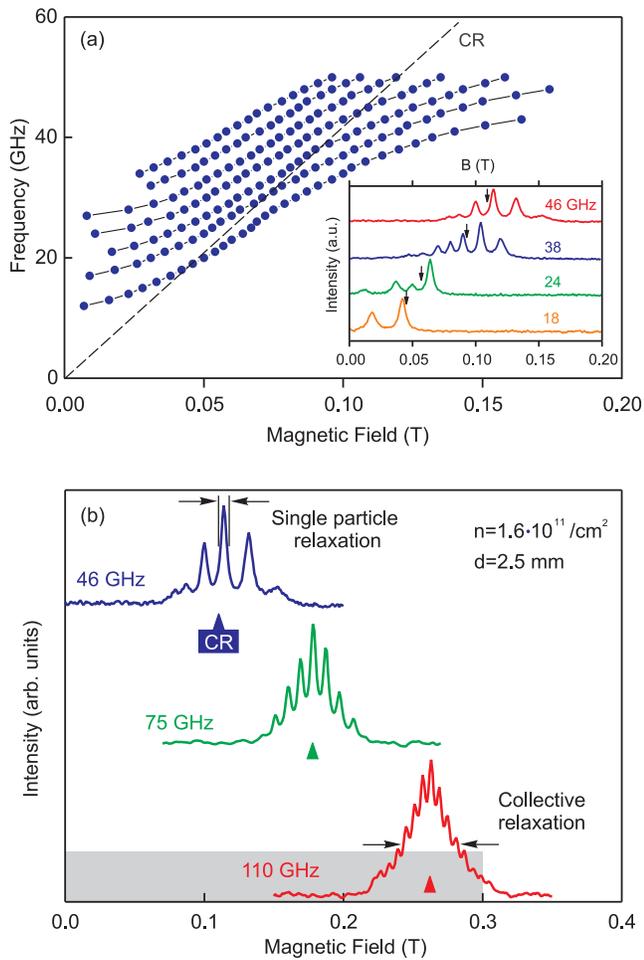}
\caption{(a) Magnetodispersion of the series of the magnetoplasma modes in $d=2.5$~mm size disk with electron density of $n_s=1.6\times 10^{11}/{\rm cm}^2$. Dashed line corresponds to the cyclotron frequency in GaAs. Inset shows magnetoplasma resonance peaks in microwave absorption intensity versus magnetic field. Arrows denote position of CR. (b) Microwave absorption as a function of the magnetic field. At all three microwave frequencies, $f = 46, 75$,~and~$110$~GHz, the broad CR line is indented by a fine structure. Arrows denote position of CR. Hatched region indicates the $B$-field range in which the CR fine structure can be resolved. Curves are offset vertically for clarity.}
\label{2}
\end{figure}

To further study the CR genesis and the interplay between the single-particle and collective relaxation processes, we used a disk with a larger diameter, $d=2.5$~mm, and an electron density of $n_s=1.6 \times 10^{11}/{\rm cm}^2$. This structure features a high retardation parameter, $A \approx 1$, that equals the ratio of  the plasmon and photon frequencies at the wavevector determined by the size of the structure. The large retardation parameter indicates hybridization between the plasmon and photon modes. This leads to more pronounced manifestation of higher magnetoplasma harmonics, together with a reduced slope of the magnetodispersion curves that cross the cyclotron line~\cite{Kukushkin:03, Kukushkin:06, Mikhailov:05}. As we will see later, this property contributes to the CR fine structure resolution. The inset of Fig.~2(a) demonstrates typical dependencies of the microwave absorption versus the magnetic field for four microwave frequencies, $f=18$,~$24$,~$38$,~and~$46$~GHz. The absorption shows a series of resonant peaks. However, none of them corresponds to excitation of CR at $B_c$ (arrows in Fig.~2(a) inset). For the frequencies under consideration ($f < 50$~GHz), the incoherent single-particle damping dominates the 2D plasma relaxation $\gamma > \Gamma_{\rm disk}$. Therefore, the observed peaks correspond simply to the dimensional magnetoplasma harmonics. Their magnetodispersion dependence is summarized in Fig.~2(a) by solid dots. Multiple magnetoplasma modes intersect the CR line owing to retardation effects~\cite{Kukushkin:03, Kukushkin:06, Mikhailov:05}.

To unveil the origin of the cyclotron resonance, we continued our research on the same sample at higher irradiation frequencies. Figure~2(b) shows the magnetic field dependence of the 2DES disk absorption intensity for microwave frequencies of $f=46$,~$75$,~and~$110$~GHz. At higher excitation frequencies, multiple magnetoplasma resonances merge with each other, and their maxima start to follow a common envelope function centered precisely at the CR field $B_c$ (arrows in Fig.~2(b)). The width of the CR envelope is much larger than the linewidth of each individual magnetoplasma resonance. Considering that the condition $d > \lambda$ is already fulfilled at $110$~GHz, we interpret the broad envelope resonance as a \textit{pure} CR with width defined by a collective superradiant relaxation $\Delta \omega = \Gamma$ (see Eqs.~3-4), whereas the magnetoplasma resonance linewidth is determined by a single-particle collisional relaxation, $\Delta \omega = 1/\tau$~\cite{Andreev:14}. Therefore, the magnetoplasma resonances are superimposed on the CR contour forming a fine structure of the resonance. The CR fine structure is resolved when the coherent superradiant 2D plasma relaxation dominates the incoherent collisional damping $\Gamma_{\rm disk} \approx \Gamma > \gamma$ (which is equivalent to $2 \pi \sigma_{xx} > 1$). From the B-field widths of the resonances in Fig.~2(b), one can obtain $\Gamma/2\pi = 7{.}4$~GHz and $\gamma/2\pi = 0{.}7$~GHz.

The CR fine structure is strongly modified as the excitation frequency increases. Although the CR envelope function remains almost unchanged, the fine structure resonances merge and almost disappear at $f=110$~GHz (Fig.~2(b)). The reason is that at such high frequencies, the spacing between the dimensional magnetoplasma resonances becomes comparable to the resonance linewidth. At frequencies of $\omega \gg \omega_p$, equation~(\ref{f1}) can be expanded in a series, $\omega \approx \omega_c + \omega_p^2/2\omega_c$. Substituting here expression for the plasma frequency from Eq.~(\ref{f2}), we obtain the frequency spacing between the $N$ and $N+1$ spectral lines of the CR fine structure:
\begin{equation}
\Delta \omega = \frac{\omega_p^2 (N=1)}{2 \omega_c} = \left( \frac{2 \pi \sigma_{xy}}{c} \right) \left( \frac{\omega_{\rm photon}}{4 \varepsilon} \right), 
\label{f3}
\end{equation} 
where $\omega_{\rm photon} = 2 \pi c/d$ is the photon frequency. The conductivity $\sigma_{xy}$ has the dimensionality of speed in Gaussian units. Therefore, the first factor in~(\ref{f3}) represents a new electrodynamic constant that can be simplified very elegantly as $2 \pi \sigma_{xy}/c = \nu \alpha$, where $\alpha = e^2/\hbar c  \approx 1/137$ is the fundamental fine-structure constant, and $\nu$ is the filling factor. Therefore, we can rewrite Eq.~(\ref{f3}) in a more concise form $\Delta \omega \sim (\nu \alpha) \, \omega_{\rm photon}$. Here we have analogy with the fine structure splitting of the spectral lines of atoms. The atomic fine structure splitting can be also reduced to a very similar expression of $\Delta E \sim (Z \alpha)^2 \, \rm{Ry}$, where $Z$ is the atomic number and $\rm{Ry}$ is the Rydberg energy constant~\cite{Landau}.

The CR fine structure is resolved whenever $\Delta \omega > \gamma$, that is,
\begin{equation}
\nu \alpha \left( \frac{\omega_{\rm photon} \tau}{4 \varepsilon} \right) > 1.
\label{f4}
\end{equation} 
For all the experiments, in which magnetoplasma resonances are observed, $\omega_{\rm photon} \gg \omega_p > 1/\tau$. Consequently, the condition $\nu > 1/\alpha$ dictates the filling factor range in which the CR fine structure is necessarily resolved. For the sample under study, $\omega_{\rm photon} \tau /4 \varepsilon \approx 6$, which limits the filling factor $\nu$ interval at which the CR fine structure can be observed to $\nu > 23$ (filled region in Fig.~2(b)). The value $\nu > 23$ ($B < 0.3$~T) agrees well with the almost complete disappearance of the CR fine structure at $f=110$~GHz.

\begin{figure}[!t]
\includegraphics[width=0.47 \textwidth]{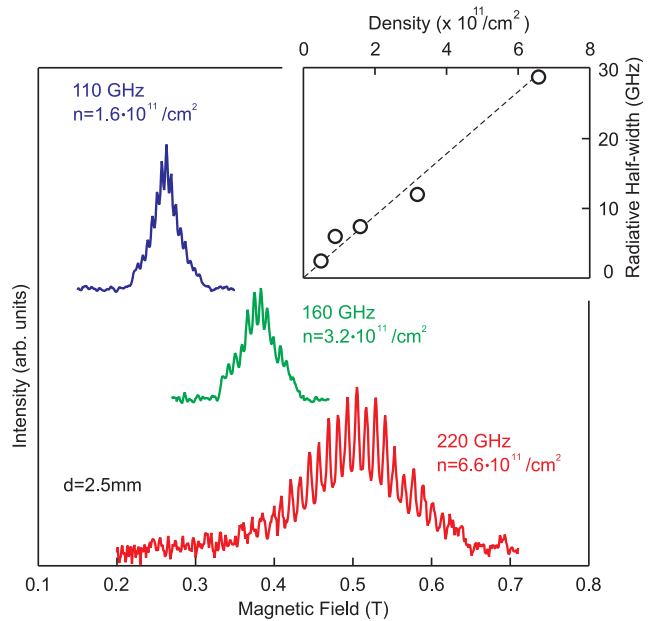}
\caption{Microwave absorption intensity as a function of magnetic field for geometrically identical samples ($d=2.5$~mm) with electron densities of $n_s = 1.6, 3.2$,~and~$6.6 \times 10^{11}/{\rm cm}^2$. Curves are offset vertically for clarity. Inset shows a plot of the CR linewidth versus the electron density. Dashed line marks the density dependence of the resonance line-width as predicted by Eq.~(\ref{gamma}).}
\label{3}
\end{figure}

We have attributed the linewidth of the CR envelope contour to the collective superradiant relaxation $\Gamma$. To further confirm our hypothesis, we repeated the experiments of Fig.~2(b) for a set of samples with different electron densities. The results for three electron densities, $n_s = 1{.}6$, $3{.}2$, and $6{.}6 \times 10^{11}/{\rm cm}^2$, are shown in Fig.~3. The cyclotron resonance clearly broadens considerably with increasing density. The inset of Fig.~3 summarizes all the superradiant linewidth points for the tested samples. The dashed line represents the theoretical prediction given by Eq.~(\ref{gamma}). There is good agreement between the theory and experiment. The quantity $\Gamma$ plays a fundamental role in light-matter interaction and can be treated as the probability of electron-photon scattering in the 2DES~\cite{Leavitt:86}.


In conclusion, we discovered that the CR originates as a \textit{pure} resonance, that does not hybridize with dimensional magnetoplasma excitations. The magnetoplasma resonances are superimposed on the CR contour forming a fine structure of the resonance. The rise of CR is observed whenever two fundamental electrodynamic relations, $2 \pi \sigma_{xx} > c$ and $2 \pi \sigma_{xy} > c$, are satisfied. The latter can be reformulated as $\nu > 1/\alpha$, where $\alpha = e^2/\hbar c \approx 1/137$ is the fine-structure constant. The experimental results agree well with the theoretical prediction. These experiments pave the way for research in the field of bizarre 2D plasma electrodynamics in the $2 \pi \sigma_{xy} > c$ and $2 \pi \sigma_{xx} > c$ regimes.   

The authors gratefully acknowledge financial support from the Russian Scientific Fund (Grant No.~14-12-00599).




\begin{thebibliography}{14}

\bibitem{CR1}
G.~Dresselhaus, A.~F.~Kip, and C.~Kittel, Phys.~Rev. {\bfseries{92}}, 827 (1953).

\bibitem{CR2}
Benjamin~Lax, H.~J.~Zeiger, R.~N.~Dexter, and E.~S.~Rosenblum, Phys.~Rev. {\bfseries{93}}, 1418 (1954).

\bibitem{CR3}
G.~Dresselhaus, A.~F.~Kip, and C.~Kittel, Phys.~Rev. {\bfseries{98}}, 368 (1955).


\bibitem{Abstreiter:74}
G.~Abstreiter, P.~Kneschaurek, J.~P.~Kotthaus, and J.~F.~Koch, Phys.~Rev.~Lett. {\bfseries{32}}, 104 (1974).

\bibitem{Allen:74}
S.~James~Allen, Jr., D.~C.~Tsui, and J.~V.~Dalton, Phys.~Rev.~Lett. {\bfseries{32}}, 107 (1974).

\bibitem{Tsui:83}
Th.~Englert, J.~C.~Maan, Ch.~Uihlein, D.~C.~Tsui, and A.~C.~Gossard, J.~Vac.~Sci.~Technol.~B {\bfseries{1}}, 427 (1983); Solid~State~Commun. {\bfseries{46}}, 545 (1983). 

\bibitem{Platzman:84}
Z.~Schlesinger, S.~J.~Allen, J.~C.~M.~Hwang, P.~M.~Platzman, and N.~Tzoar, Phys.~Rev.~B 
{\bfseries{30}}, 435(R) (1984).

\bibitem{Zudov:01}
M.~A.~Zudov, R.~R.~Du, J.~A.~Simmons, and J.~L.~Reno, Phys.~Rev.~B {\bfseries{64}}, 201311(R) (2001).

\bibitem{Kukushkin:02}
I.~V.~Kukushkin, J.~H.~Smet, K.~von~Klitzing, and W.~Wegscheider, Nature {\bfseries{415}}, 409 (2002).

\bibitem{ZnO}
Y.~Kasahara, Y.~Oshima, J.~Falson, Y.~Kozuka, A.~Tsukazaki, M.~Kawasaki, and Y.~Iwasa, Phys.~Rev.~Lett. {\bfseries{109}}, 246401 (2012).

\bibitem{Vankov:15}
V.~E.~Kozlov, A.~B.~Van'kov, S.~I.~Gubarev, I.~V.~Kukushkin, V.~V.~Solovyev, J.~Falson, D.~Maryenko, Y.~Kozuka, A.~Tsukazaki, M.~Kawasaki, and J.~H.~Smet, Phys.~Rev.~B {\bf 91}, 085304 (2015).

\bibitem{Shayegan:93}
T.~S.~Lay, J.~J.~Heremans, Y.~W.~Suen, M.~B.~Santos, K.~Hirakawa, M.~Shayegan, and A.~Zrenner, Appl.~Phys.~Lett. {\bf 62}, 3120 (1993).

\bibitem{Muravev:05}
V.~M.~Muravev, A.~R.~Khisameeva, V.~N.~Belyanin, I.~V.~Kukushkin, L.~Tiemann, C.~Reichl, W.~Dietsche, and W.~Wegscheider, Phys.~Rev.~B {\bf 92}, 041303(R) (2015).

\bibitem{Allen:83}
S.~J.~Allen and H.~L.~St\"{o}rmer and J.~C.~M.~Hwang, Phys.~Rev.~B {\bf 28}, 8 (1983).

\bibitem{Heitmann:88}
T.~Demel, D.~Heitmann, P.~Grambow, and K.~Ploog, Phys.~Rev.~B {\bf 38}, 12732 (1988).

\bibitem{Shikin:92}
V.~Shikin, T.~Demel, and D.~Heitmann, Phys.~Rev.~B {\bf 46}, 3971 (1992).

\bibitem{Stern}
F.~Stern, Phys.~Rev.~Lett. {\bf 18}, 546 (1967).

\bibitem{Ashkinadze}
B.~M.~Ashkinadze, E.~Linder, E.~Cohen, and Arza Ron, Phys.~Stat.~Sol. {\bf 164}, 231 (1997).

\bibitem{Gubarev}
M.~Y.~Akimov, I.~V.~Kukushkin, S.~I.~Gubarev, S.~V.~Tovstonog, J.~Smet, K.~von~Klitzing, and W.~Wegscheider, Zh.~Eksp.~Teor.~Fiz. {\bf 72}, 662 (2000)[JETP Lett. {\bf 72}, 460 (2000)].

\bibitem{Allen:83}
S.~J.~Allen, Jr., H.~L.~St\"{o}rmer, and J.~C.~M.~Hwang, Phys.~Rev.~B {\bfseries{28}}, 4875(R) (1983).

\bibitem{Vasiliadou:93}
E.~Vasiliadou, G.~M\"{u}ller, D.~Heitmann, D.~Weiss, K.~v.~Klitzing, H.~Nickel, W.~Schlapp, and R.~L\"{o}sch, Phys.~Rev.~B {\bfseries{48}}, 17145 (1993).

\bibitem{Andreev:14}
I.~V.~Andreev, V.~M.~Muravev, V.~N.~Belyanin, and I.~V.~Kukushkin, Appl.~Phys.~Lett. {\bfseries{105}}, 202106 (2014).

\bibitem{Quinn:76}
W.~Chiu, T.~K.~Lee, and J.~J.~Quinn, Surf.~Sci. {\bfseries{58}}, 182 (1976).

\bibitem{Mikhailov:96}
S.~A.~Mikhailov, Phys.~Rev.~B {\bfseries{54}}, 10335 (1996).

\bibitem{Matov:96}
O.~R.~Matov, O.~F.~Meshkov, O.~V.~Polishchuk, and V.~V.~Popov, JETP {\bfseries{82}}, 471 (1996).

\bibitem{Mikhailov:04}
S.~A.~Mikhailov, Phys.~Rev.~B {\bfseries{70}}, 165311 (2004).

\bibitem{Zudov:14}
Qi~Zhang, T.~Arikawa, E.~Kato, J.~L.~Reno, W.~Pan, J.~D.~Watson, M.~J.~Manfra, M.~A.~Zudov, M.~Tokman, M.~Erukhimova, A.~Belyanin, and J.~Kono, Phys.~Rev.~Lett. {\bfseries{113}}, 047601 (2014).

\bibitem{Leavitt:86}
R.~P.~Leavitt and J.~W.~Little, Phys.~Rev.~B {\bfseries{34}}, 2450 (1986).

\bibitem{Kukushkin:03}
I.~V.~Kukushkin, J.~H.~Smet, S.~A.~Mikhailov, D.~V.~Kulakovskii, K.~von~Klitzing, and W.~Wegscheider,
Phys.~Rev.~Lett. {\bfseries{90}}, 156801 (2003).

\bibitem{Kukushkin:06}
I.~V.~Kukushkin, V.~M.~Muravev, J.~H.~Smet, M.~Hauser, W.~Dietsche, and K.~von~Klitzing,
Phys.~Rev.~B {\bfseries{73}}, 113310 (2006).

\bibitem{Mikhailov:05}
S.~A.~Mikhailov, N.~A.~Savostianova, Phys.~Rev.~B {\bf 71}, 035320 (2005).

\bibitem{Landau}
L.~D.~Landau, L.~M.~Lifshitz, \textit{"Quantum Mechanics: Non-Relativistic Theory"}, Elseiver (2003).


\end{thebibliography}
\end{document}


\title{Supplementary Material for\\ ``Fine Structure of Cyclotron Resonance in a Two-Dimensional Electron System''}
\author{V.~M.~Muravev$^{a}$, I.~V.~Andreev$^{a}$, S.~I.~Gubarev$^{a}$, V.~N.~Belyanin$^{a,b}$, I.~V.~Kukushkin$^{a}$}
\affiliation{$^a$ Institute of Solid State Physics RAS, Chernogolovka 142432, Russia\\
$^b$ Moscow Institute of Physics and Technology, Dolgoprudny 141700, Russia\\ }

\date{\today}\maketitle

\section{\textrm{I}. OPTICAL TECHNIQUE FOR DETECTION OF PLASMA RESONANCES IN 2DES}

The optical technique used in our experiments to measure the microwave absorption in 2DES is based on the high sensitivity of the 2D electron luminescence spectrum to 2DES resonant heating~\cite{Ashkinadze, Gubarev}. Basically, this technique allows us to detect 2DES heating due to resonance plasmon excitation. One unique aspect of the optical technique is that it is non-invasive. No metallic gates or contacts, which inevitably disturb the plasmon field, are needed near the structure. This property makes the optical detection one of the most accurate methods for studying plasma excitations in semiconductors. 

\begin{figure}[!h]
\includegraphics[width=0.74 \textwidth]{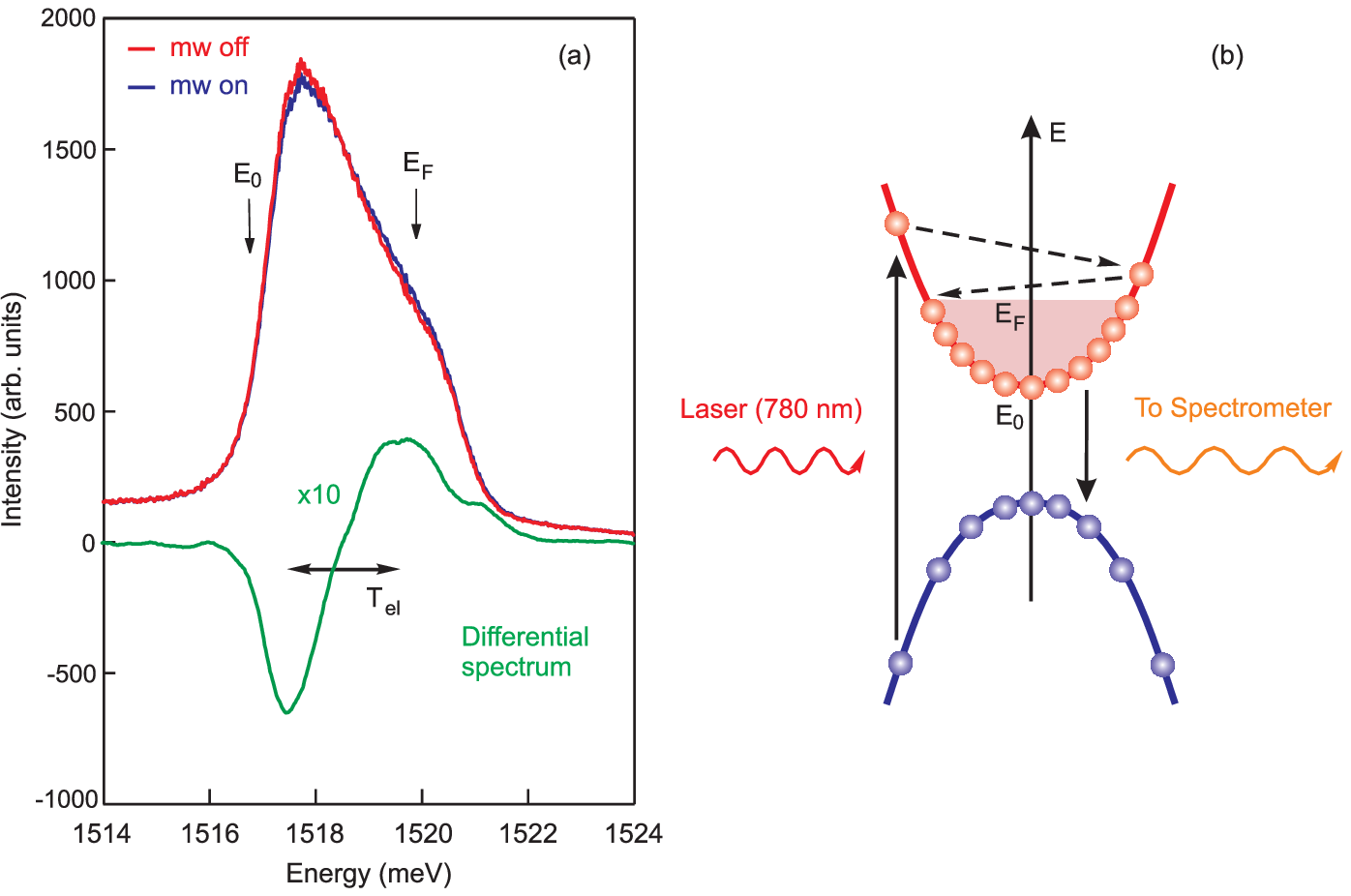}
\caption{((a)Typical luminescence spectrum with (blue line) and without (red line) a $50$~$\mu$W microwave excitation at $22.5$~GHz obtained in a disk-shaped 2DES with a diameter of $1$~mm and carrier density $0{.}9 \times 11^{10}/{\rm cm}^2$ at a magnetic field $B = 60$~mT. In the vicinity of the Fermi energy $E_{\rm F}$, the spectrum is affected significantly under resonant microwave excitation due to heating. The bottom line (green) represents the differential luminescence spectrum. The width of the differential spectrum reflects the increased electron temperature $T_{\rm el}$. The integration of its absolute value across the entire spectral range yields the microwave absorption amplitude recorded in our experiments. (b) Schematic energy diagram illustrating optical transitions that take place in 2DES under laser irradiation.
}
\label{Optics}
\end{figure}

To study 2DES luminescence spectra, a sample was illuminated with a 780-nm stabilized CW semiconductor laser diode (Thorlabs CPS192) through an optical quartz fiber passed into the waveguide. Diode was feeded by highly-stable DC source (Yokogawa 7651) and placed in an air-conditioned lab room with constant air temperature so its power stability was better than $0.1 \%$. Light from the sample was collected by the same fiber, and then luminescence spectra were recorded using a liquid nitrogen cooled charge-coupled device camera (Princeton Instruments LN/CCD-1340/100-EHR/1) and a double monochromator spectrometer (Coderg PHO) with a spectral resolution of $0.03$~meV.  A laser with wavelength of $\lambda=780$ was selected so that the energy of the laser photon lies in a range between the energy gaps in quantum well and barrier: $E_{g}^{\rm GaAs} < h \lambda / c < E_{g}^{\rm Al_{0.33}Ga_{0.67}As}$. This laser generates electron-hole pairs in the GaAs quantum well, but not in the AlGaAs barrier and spacer. Laser power was kept constant during the experiment and typically did not exceed $0.1$~mW. For such power level steady-state density of holes generated by laser illumination in the quantum well is about $n_h = 10^7 - 10^8 $cm$^{-2}$, and electron density in the quantum well remains almost unchanged. This has been proven by independency of plasma frequency on excitation laser power. 

The shape of 2D electrons recombinant luminescence spectrum is given by a convolution of 2D electron and hole distribution functions $f_e, f_h$ and densities of states $D_e, D_h$, and the square of the absolute value of transition matrix element $W_{cv}=|P_{cv}|^2$ : 
\begin{equation}
I(\omega) = \int_0^{\infty}W_{cv}(E, \hbar\omega-E)D_e(E)D_h(\hbar\omega-E)f_e(E)f_h(\hbar\omega-E)dE.
\end{equation}
In our measurements an effect of Landau quantization on $D_e$ and $D_h$ appears in the luminescence spectrum only in magnetic fieds $B>400$~mT, and can be neglected in smaller magnetic fields. Since electron distribution function $f_e$ obeys Fermi-Dirac statistics, and the transition matrix element may be assumed to be constant, luminescence spectrum $I(\omega)$ directly represents single-particle hole distribution function $f_h$, which obeys Boltzmann statistics and therefore is sensitive even to small 2DES heating due to the microwave absorption. 

In our experiment we studied the difference between 2DES luminescence spectra with and without microwave irradiation of the sample at fixed magnetic field (Fig.~\ref{Optics}S). We recorded the differential luminescence spectrum and used the integral of the absolute value of the differential spectrum as a measure of the microwave absorption intensity of the sample. Next, another magnetic field is chosen, and the above procedure is repeated to obtain another data point. And so on so forth, before the whole absorption curve is obtained (see Fig.~1(a)).

\section{\textrm{II}. MICROWAVE POWER DEPENDENCY OF PLASMA RESONANCE ABSORPTION}

\begin{figure}[!t]
\includegraphics[width=0.6 \textwidth]{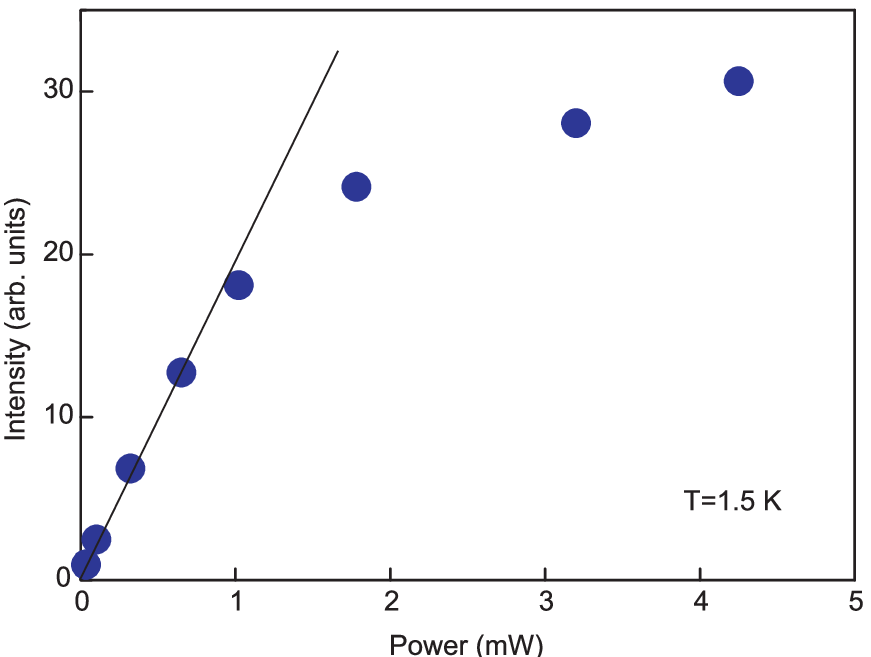}
\caption{Dependency of the 2DES microwave absorption intensity as measured by the optical technique versus microwave power at the input of the waveguide ($f=20$~GHz).}
\label{Power}
\end{figure}

Fig.~\ref{Power}S shows a typical dependency of the 2DES microwave absorption intensity measured by the optical technique versus microwave power at the input of the waveguide ($f=20$~GHz). The sample was mounted in the center of the $8 \times 16 \, {\rm mm}^2$ rectangular waveguide. At frequency $f=20$~GHz the sample (disk of $d=1$~mm and $n_s=0{.}9 \times 10^{11}/{\rm cm}^2$ ) is located at the antinode of standing $\rm{TE}_{10}$ electromagnetic wave. Therefore, the coupling of microwave radiation to the sample at this frequency is most effective. The data show that optically measured microwave absorption intensity does linearly depend on the microwave power absorbed by the sample for small input power levels of $P < 1$~mW. 

In our experiments, microwave input power did not exceed $0.5$~mW for the frequency range $12-70$~GHz, and $0.1$~mW for the frequency range $70-250$~GHz. As such, we always work in the linear response regime. 

\section{\textrm{III}. ANALYSIS OF 2D PLASMON MAGNETODISPERSION IN A DISK SHAPED SAMPLES}

\begin{figure}[!h]
\includegraphics[width=0.6 \textwidth]{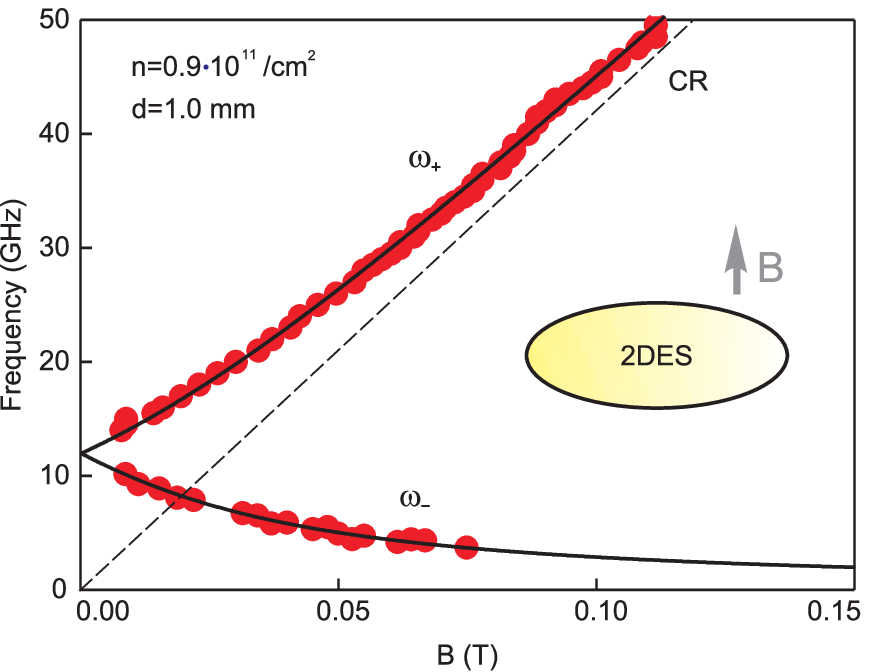}
\caption{Resonance magnetic field position for $n_s = 0{.}9 \times 10^{11}/{\rm cm}^2$ and $d=1$~mm as a function of incident microwave frequency. The solid lines represent the theoretical dependence of the hybrid dimensional magnetoplasma-cyclotron resonance according to Equation~(\ref{B}). The dashed line corresponds to the pure cyclotron mode.}
\label{Dispersion}
\end{figure}

Figure~\ref{Dispersion}S shows the magnetodispersion of 2D plasma excitation for a single disk having a diameter of $d=1$~mm. The data were obtained at an electron density of $0{.}9 \times 10^{11}/{\rm cm}^2$. The magnetodispersion has two branches. The low-frequency branch $\omega_{-}$ corresponds to an edge magnetoplasmon (EMP) propagating along the edge of the disk. This is a mode with anomalously weak attenuation that propagates in a narrow strip near the edge of the 2DES~\cite{Volkov:88, Allen:83}. The high-frequency branch $\omega_{+}$ has a positive magnetodispersion. The plasma excitation spectrum in a 2DES with a disk shaped geometry can be described using the dipole approximation as~\cite{Dahl, Heitmann}:

\begin{equation}
\omega_{\pm} = \pm \frac{\omega_c}{2} + \sqrt{\omega_p^2 + \left(  \frac{\omega_c}{2} \right)^2},
\label{B}
\end{equation}

where $\omega_p$ is plasma frequency in the zero magnetic field~\cite{Stern:67}:

\begin{equation}     
\omega_p^2=\frac{2 \pi n_s e^2}{m^{\ast} \varepsilon} q.
\label{F}
\end{equation}      

Here $\varepsilon^{\ast}=(\varepsilon_{\rm GaAs}+1)/2$ is the effective dielectric permittivity of the surrounding medium and $q=2{.}4/d$ is the wave vector for the disk geometry~\cite{Fetter:86, Kukushkin:03}. Solid lines in Fig.~\ref{Dispersion}S represent best fit for experimental data by Equation~(\ref{B}). As a result, we get $f_p=\omega_p/2 \pi = (12 \pm 2)$~GHz. This value of experimental plasma frequency aligns acceptably with the theoretical prediction from Equation~(\ref{F}) $f_p=13.5$~GHz. The minor discrepancy can likely be ascribed to the inaccurate description
of the dielectric environment surrounding the 2DES.    

